\documentclass{llncs}

\newcommand{\systemname}[1]{\textsf{#1}}
\newcommand{\tipi}[0]{\systemname{Tipi}}

\newcommand{\vampire}[0]{\systemname{Vampire}}
\newcommand{\eprover}[0]{\systemname{E}}

\newcommand{\provernine}[0]{\systemname{Prover9}}
\newcommand{\hilbertnine}[0]{\systemname{Hilbert9}}
\newcommand{\pnineloop}[0]{\systemname{p9loop}}
\newcommand{\prooftrans}[0]{\systemname{Prooftrans}}

\newcommand{\sad}[0]{\systemname{SAD}}

\usepackage{hyperref}
\newcommand{\tptpproblemurl}[1]{http://www.cs.miami.edu/~tptp/cgi-bin/SeeTPTP?Category=Problems\&Domain=GRA&File={#1}.p}
\newcommand{\tptpproblemlink}[1]{\href{\tptpproblemurl{#1}}{\texttt{#1}}}
\newcommand{\tptpaxiomurl}[1]{http://www.cs.miami.edu/~tptp/cgi-bin/SeeTPTP?Category=Axioms\&File={#1}.ax}

\usepackage{mathabx}

\usepackage{mathptmx}
\usepackage{hyperref}
\title{Proof identity for mere mortals }
\titlerunning{Proof identity}
\author{Jesse Alama\inst{1}\thanks{The author was partially supported by FWF grant P25417-G15 (LOGFRADIG) as well as by the Portuguese Science Foundation (FCT) project \emph{The notion of mathematical proof} (PTDC/MHC-FIL/5363/2012).}}
\authorrunning{Alama}
\institute{Theory and Logic Group{\\}Technical University of Vienna{\\}\email{alama@logic.at}}

\begin{document}
\maketitle
\begin{abstract}
The proof identity problem asks: When are two proofs the same?  The question naturally occurs when one reflects on mathematical practice.  The problem understandably can be seen as a challenge for mathematical logic, and indeed various perspectives on the problem can be found in the proof theory literature.  From the proof theory perspective, the challenge is met by laying down new calculi that eliminate ``bureaucracy''; techniques such as normalization and cut-elimination, as well as proof compression, are employed.  In this note a new perspective on the proof identity problem is outlined.  The new approach employs the concepts and tools of automated theorem proving and complements the rather more theoretical perspectives coming from pure proof theory.  The practical approach is illustrated with experiments coming from the TPTP Problem Library.
\end{abstract}
\section{Introduction}\label{sec:introduction}
The proof identity problem asks: \emph{When are two proofs the same?}  The question has undoubtedly occurred to anyone who has reflected on mathematical practice.  Generally, having multiple proofs of a result is generally regarded as valuable.  (For a philosophical perspective, see~\cite{hersh1997prove}.)  Some results in mathematics receive considerable scrutiny, in the sense that they are given many different proofs.  Consider, for instance, Carl Gauss' many proofs of the quadratic reciprocity theorem in number theory~\cite{lemmermeyer2000reciprocity}\footnote{At \url{http://www.rzuser.uni-heidelberg.de/~hb3/rchrono.html} H.~Lemmermeyer maintains a list of proofs of the quadratic reciprocity law.  At the time of writing, Lemmermeyer's list has 240 entries.}, or the diversity of proofs of the Pythagorean theorem~\footnote{The mathematics site Cut the Knot maintains 99 proofs: \url{http://www.cut-the-knot.org/pythagoras/index.shtml}}.  If one teaches mathematics, the experience of grading students' work is often that one finds that successful solutions to mathematical problems can be clusters: some students seem to solve the problem one way, others another way.  A new proof may even be a strong desideratum, such as the case of the four-color theorem, which for a few decades was widely regarded as correct but proved in a discomfiting way~\cite{gonthier2008four}.

But what is a ``way of proving''?  The question is perhaps not well-posed.  Mathematical logic gives us some tools for making the problem somewhat more concrete (e.g., the completeness theorem for first-order logic, the wide catalog of available calculi for formally proving theorems).  But the underlying notion of identity of proof, even in a formal setting, seems to be somewhat slippery; the proof identity problem might is perhaps just a proxy for the whole field of proof theory.

Perhaps the problem can be addressed indirectly: when is one proof simpler than another?  One might perhaps be interested in the simplest proof of a theorem under certain conditions.  Here we can take some comfort in the presence of company.  Hilbert's (previously unknown) 24th problem~\cite{thiele2003hilbert} addresses precisely this issue.  Already in 1900 Hilbert was thinking about the difficulty:

\begin{quotation}
The 24th problem in my Paris lecture [the presentation in 1900 at the International Congress of Mathematicians where Hilbert's famous problem list was given] was to be: Criteria of simplicity, or proof of the greatest simplicity of certain proofs. Develop a theory of the method of proof in mathematics in general. Under a given set of conditions there can be but one simplest proof. Quite generally, if there are two proofs for a theorem, you must keep going until you have derived each from the other, or until it becomes quite evident what variant conditions (and aids) have been used in the two proofs. Given two routes, it is not right to take either of these two or to look for a third; it is necessary to investigate the area lying between the two routes.
\end{quotation}

Hilbert can be considered one of the central figures in the history of proof theory; his student Gentzen gave us natural deduction and sequent calculus, two formalism that surely cannot be ignored in any discussion of the proof identity problem.  See~\cite{prawitz2000ideas} for a readable perspective on some central results in proof theory such as normalization in natural deduction and cut-elimination for sequent calculi.  In recent years there has been an interest in pushing cut-elimination, which customarily is understood as not applying to non-logical axiom systems (in other words, in general cut-elimination fails as soon as one postulates a non-logical axiom), to the analysis of axiom systems~\cite{negri2008structural,negri2013proof}.

We are thus interested in the problem, loosely understood, of what it means for two proofs to be the same.  Hilbert's suggested approach, maximizing simplicity (however understood), might be followed.  Can we make these terms more precise?

At first blush, one immediately faces a difficulty: the \emph{Multiplicity of Proof} problem:

\begin{quotation}
If there is one proof of a theorem, then there is (probably) another.
\end{quotation}

{\noindent}When one looks at proofs through a formal lens, one generally encounters a somewhat sharper phenomenon, which we might call the \emph{Infinitude of Proof} problem:

\begin{quotation}
If there is one proof of a theorem, then there are (in fact) infinitely many.
\end{quotation}

{\noindent}Thus, consider a Hilbert-style calculus, a formula $\phi$ and a set $X$ of axioms, and a derivation $d$ (a certain kind of sequence) of $\phi$ from $X$.  The derivation $d$ witnesses the derivability (which we identify with provability) of $\phi$ from $X$.  Monotonicity allows us to add new, unused premises to $X$, simply prepending them to $d$.  If one objects to adding new proper axioms, one can simply take new logical axioms (or instances thereof) and prepend them to $d$.  We thereby obtain, apparently, a new derivation of $\phi$.  In $d$ we may be able to permute certain terms, thereby yielding another (?) derivation of $\phi$ from $X$.  No end seems to be in sight.

Combating such multiplicity is a genuine problem.  Such ``attacks'' on the notion of ``proof of $\phi$ from premises $X$'' can, in part, be addressed through trivial requirements such as:

\begin{itemize}
\item ensuring that all members of $X$ are actually used somewhere in the proof (no spurious premises)
\item ensuring that from the conclusion there is a path back to all axioms (a stronger form of the first condition)
\item permitting no duplication (it is not necessary to have two
  occurrences of one and the same formula in a derivation)
\item identifying derivations that are the same up to a permutation
\end{itemize}

{\noindent}But such structural conditions are just the tip of the iceberg.  One can derive $\phi$ from $\phi \wedge \phi$, and from $\phi \vee \vee$; what if the latter are postulated as axioms?  Such conundrums presumably ought to be addressed.  One could object that no proof depends on such tricks.  But if we want a robust notion of redundancy, we presumably have to address these kinds of obstacles.

To combat such maddening multiplicity, one naturally wants to define and employ a notion of redundancy.  Two derivations could be identified if they are the same up to the notion of redundancy.  One can employ various syntactic and semantic notions to help define such notions of redundancy.

In two other models of mathematical proof, sequent calculus and natural deduction, the situation seems to be no different.  The ``if there is one, there are infinitely many'' phenomenon also seems to hold sway.  One can define notions of normalization.  But then one is boxed in from another direction: if we employ normalization of derivations, how do we avoid the problem that theorems are always proved in one way?  This also seems intuitively unacceptable.  We don't want to rule out the multiplicity of proofs (or at least, the potential multiplicity of proofs) by identifying all proofs.

Going the other way around, one can design normalization procedures that identify proofs.  But then one encounters the notion of normalization in proof theory.  Here, the interest is in defining suitable reduction relations among derivations and proving that an arbitrary derivation can be put into a normal form.  Even further, one finds interest in reducing derivations to a unique normal form (strong normalization).  Such an approach witnesses the following \emph{Canonicity Principle}:

\begin{quotation}
Every theorem has a canonical proof.
\end{quotation}

On the theoretical side, there is a complexity problem: normalization and cut-elimination often lead to tremendous blowup.  Starting with a ``moderately'' small natural deduction derivation, one may very well find that its normal form is so large as to be unworkable:

\begin{quotation}
Faced with two concrete proofs in mathematics---for example, two proofs of the Theorem of Pythagoras, or something more involved---it could seem pretty hopeless to try to decide whether they are identical just armed with the Normalization Conjecture or the Generality Conjecture. But this hopelessness might just be the hopelessness of formalization. We are overwhelmed not by complicated principles, but by sheer quantity.~\cite{Dos03}
\end{quotation}

{\noindent}(The two conjectures mentioned here, the Normalization Conjecture and the Generality Conjecture, are proposals for addressing the proof identity problem.) Indeed, normalized proofs are often quite large.  (Nonetheless, the situation is not always hopeless~\cite{baaz2011methods}.)  One can design alternative calculi with the hope of cutting down on the tedious ``bureaucracy'' that generally accompanies any non-trivial formalization effort~\cite{Gugl:05:The-Prob:nu}.

Thus it seems that ``theory'', that is, pure mathematical logic (proof theory), offers few genuinely practical solutions to the proof identity problem.  Sure, one can implement these algorithms, but if one is interested in mathematics (as opposed to just pure logic[s]), we need to tamper our expectations.   We seek tools that we can apply to many examples, we want to get our hands on normalized derivations, but it seems that the paths down that road are prohibitively expensive.  Knuth put it best:

\begin{quotation}
Of course, computers are not infinitely fast, and our expectations have become inflated even faster than our computational capabilities.  We are forced to realize that there are limits beyond which we cannot go.  The numbers we can deal with a not only finite, they are very finite, and we do not have the time or space to solve certain problems even with the aid of the fastest computers.~\cite{knuth1976coping}
\end{quotation}


In this paper another perspective is taken.  Towards addressing Hilbert's 24th problem, we eschew the solutions coming from pure proof theory and turn toward automated theorem provers.  Our perspective is not entirely new; L.~Wos, for instance, has discussed Hilbert's 24th problem in the context of automated theorem proving for some time~\cite{wos2003missing,thiele2002hilbert}

\begin{quotation}
As we understand more fully the range of mechanical mathematics, we get a clearer view of the relation between complexity and conceptual difficulty in mathematics, since we would probably wish to say that mechanizable pieces, even when highly complex, are conceptually easy. When alternative proofs are fully mechanizable, we obtain also a quantitative measure of the simplicity of mathematical proofs, to supplement our vaguer but richer intuitive concept of simplicity. With the increasing power to formalize and mechanize, we are freed from tedious details and can more effectively survey the content and conceptual core of a mathematical proof.~\cite{wang1963mechanical}
\end{quotation}

{\noindent}Wang's programmatic comment was made in 1963, when the field of automated theorem was very young and when computers were also fantastically weaker than they are today.  Thanks to decades of work and early efforts to develop practical solutions in this field, we can safely say that Wang's programmatic comment is more approachable today than it was when he made it.

An important contribution of mathematical logic the design of various metrics for formal proofs, such as Herbrand complexity~\cite{leitsch1997resolution} and their relation to other metrics, such as cut-complexity.  Intuitively, Herbrand complexity provides a calculus-independent measure of the complexity of a theorem, and certain proofs of the theorem can come more or less close to witnessing this minimal complexity.

So if we do not accept every theorem has exactly one proof, but we want to turn away from the infinity of proofs that mathematical logic tells us always exist, then how can we proceed?  What is the \emph{practical} proof identity problem?  Or: what is an appropriate notion of proof such that one can tackle the proof identity problem in a practical way?  We want to investigate theorems.  The ``space'' of theorems shouldn't be infinite, but it shouldn't be too big, either.  Mathematical logic does not give us the answer we want.  Where can we turn?


\section{Measurements on refutations}\label{sec:counting}
If one is interested in the proof identity problem, one might take notice of the following extract from the {\provernine} manual:

\begin{quotation}
\textbf{Multiple Proofs}

\verb+assign(max_proofs, n).  % default n=1, range [-1 .. INT_MAX]+

This parameter tells Prover9 to stop searching when the $n$-th proof has been found.
\end{quotation}

From a theoretical standpoint, a proof theorist interested in the proof identity problem ought to be stunned by such a statement.  The immediate question is: what is the notion of proof employed by {\provernine}?  How can it distinguish between different proofs?  Is it at all practical to enumerate different proofs?  By what criteria would one know when to stop the enumeration?  Aren't there always infinitely many proofs, if there is even one proof?

The answer is that {\provernine} uses a standard set-of-support search strategy in (essentially) a resolution calculus~\cite{leitsch1997resolution}.  From a theoretical view, the calculus is quite simple.  As has been standard for decades, to derive $\phi$ from axioms $X$, one considers the clause logic analogue of $X \cup \{ \neg \phi \}$ and attempts to derive a contradiction.

But from the theoretical point of view, this is not yet an answer; the proof identity problem is not solved by~(1)~switching to clause logic and~(2)~reducing the arbitrary proof-finding problem to the problem of finding a refutation. In this setting, one can replay the intuitive arguments given in the previous section for why---in the absence of special assumptions---there ought to exist infinitely resolution proofs.

The notion of proof at play behind {\provernine} and similar systems is, however, informed by decades of experience dealing with \emph{notions of redundancy}~\cite{DBLP:books/el/RV01/BachmairG01}.  The point is that various ``rabbit holes'' that theoretically exist can be avoided in this setting~\cite{DBLP:books/el/RV01/Weidenbach01}.  The goal is often to find a refutation by generating clauses (that is, drawing inferences).  In my automated theorem proving systems based on resolution or related clause-based calculi, by default, when a contradiction (the empty clause) is derived, the search terminates.  The principal problem for automated theorem proving is to determine the existence of at least one proof.  Derivability is already a very hard problem, confirmed by decades of experience of many researchers.

But if we can find one proof, why not keep going?  Just keep deriving inferences until no more inferences can be drawn.  Every time the empty clause is derived, simply output the path by which we got to this particular occurrence of the empty clause.

\section{Experiments}\label{sec:experiments}
We now proceed to consider a variety of specific test cases on which to explore Hilbert's 24th problem and the attendant proof identity problem.  We consider four proof-finding problems coming from the TPTP library~\cite{DBLP:journals/jar/Sutcliffe09}.  To facilitate our exploration, we implemented a custom tool that we call {\hilbertnine}.  {\hilbertnine} is not a new proof search tool; it is (like {\pnineloop}~\cite{kinyon2013loops}) essentially just a wrapper around {\provernine} that accomplishes, in one fell swoop, what can easily be done by hand with the {\provernine} suite.

\subsection{Partition of monoids}\label{sec:partition}
We now consider a problem about partitioning monoids (\tptpproblemlink{ALG011-1}).  We are to show that, if a monoid $M$ is partitioned into (nonempty) subsets $C$ and $D$, it is not possible that both $C \times C \subseteq D$ and $D \times D \subseteq C$.

For this problem, {\provernine} in its automatic mode is able to produce $374$ proofs before running out of possible inferences.  This number is obviously quite large, and it is quite unlikely that these proofs are equal interest.  But we find that there are $2$ proofs having the minimal length $20$.  (The longest proof has length 44.)  If we look at these two proofs, one features conclusions (clauses):

\begin{itemize}
\item \verb+c(f(A,B)) | -d(A) | c(B)+
\item \verb+c(f(a1,f(a1,A))) | c(A)+
\end{itemize}

where the binary function symbol \verb+f+ is the multiplication of the monoid, the unary predicates \verb+c+ and \verb+d+ represent the partition of the monoid, and the constants \verb+a1+ and \verb+a2+ are witnesses to the nonemptiness of the partitions (Prolog-style variables are being used, so \verb+A+ and \verb+B+ are variables).  The other minimal length derivation draws the inferences:

\begin{itemize}
\item \verb+c(f(a1,f(a1,A))) | -d(A)+
\item \verb+-d(f(a1,a2))+
\end{itemize}

All other clauses in the two derivations are shared.  The question we now face is whether these two ``combinatorially distinct'' derivations are materially different.  Notice that the second features a (negative) literal, but the distinctive two inferences drawn in the first proof are both non-literals.  This particular problem has the effect that in it \verb+c+ and \verb+d+ are opposites of one another.  Thus, the second distinctive clause of the first proof is, in effect, the same as the first distinctive clause of the first proof.  With this understanding, these two proofs are therefore probably the same and we are have one minimal (length) proof.
\subsection{Tarski-Knaster fixed-point theorem}\label{sec:tk}
\tptpproblemlink{LAT381+1} is a formalization of a Tarski-Knaster result, asserting the uniqueness of suprema in the lattice of subsets of a set.  The result is generated from another development by the {\sad} system~\cite{VLP07}.

For this problem, 8 proofs can be fairly quickly found.  There may a 9th proof; it appears that hundreds of thousands of given clauses need to be generated before a ninth proof is found.

Among the 8 proofs, the minimal length is 30, and there are 2 proofs having this length and the maximum length is 45 (and there are 4 such proofs).  They differ only in the way they derive the empty clause.  In other words, the two proofs are identical up to their final steps.  If one inspects the precise premises involved in different steps (that is, the premises of the final rule application(s) of the two proofs), they are exactly the same.  A difference can be detected, then, only if one were to use, say, {\prooftrans}'s expand command to rewrite these proofs in an even more explicit manner than {\provernine}'s proof objects.  One could argue, then, that there is in fact only one shortest proof here.

One can, however, improve on the result by eliminating redundant premises.  As mentioned, this particular problem is the result of translating an interactively developed proof.  A typical feature of such translations is that numerous additional premises are present in the problem.  This does not reflect laziness or disregard for the usual preference in mathematics and logic for parsimonious assumptions; rather, it reflects in the first place the desire to maintain completeness when one translates a development in one logic (or language) into the language of pure FOL (see, e.g.,~\cite{DBLP:conf/birthday/UrbanV13,DBLP:conf/lpar/AlamaKU12}).  In short, the translation problem is in focus, not the minimality of the translated premises (which raises its own issues, e.g., the non-uniqueness of a minimal set).

An inspection of \tptpproblemlink{LAT381+1} with the {\tipi} premise-minimization tool~\cite{alama2012tipi} finds that, of the 16 available premises, a (semantically minimal) subset of 6 can be found that suffice to derive the conjecture.  When {\provernine} is applied to the minimized problem, one obtains a sharper result.  One can now unqualifiedly show that only $5$ proofs exist (as opposed to making the qualified claim that only 8 could be found).  The ``space'' of possible proof lengths, however, is shifted while at the same time narrower.  Thus, the minimum length of the proofs increases but the maximum length decreases: the shortest proof now has length 28 (and there are 3 such proofs), whereas the longest proof has length 29 (and there are 2 such proofs).  27 clauses are held common to all 5 derivations, which gives reason to believe that the differences between the 5 possible proofs are likely quite small.  Our results are reminiscent of the approach taken when following what Wos calls cramming~\cite{wos2003cramming}.


\subsection{Chinese remainder theorem}\label{sec:chinese}
The problem (\tptpproblemlink{RNG126+1}) is to show a kind of Chinese remainder theorem for rings.  As with the Tarski-Knaster-like theorem discussed in Section~\ref{sec:tk}, this problem was generated by the {\sad} system.  Similarly, the problem has a few redundant premises whose removal can give us a different view of the problem.  Thus, the problem has 47 premises; a minimal version has only 9 premises.

When working with the original version, we seem to have to impose some limits that might affect the completeness of our method.  This is understandable in light of the fact that this is a rather hard problem (its rating in the 6.0.0 version of the TPTP library is 0.70~\cite{DBLP:journals/aicom/Sutcliffe13}, which means that most systems were not able to solve it within the limits of the CASC competition).

If we turn to the minimized version, we are able to relax the limits and let {\provernine} explore the search space with fewer constraints.  In the original problem, we found only one proof (presumably there are more, but the nature of the problem and limits of time and space prevent us from doing a complete investigation).  Interestingly, only two proofs of the minimized problem are available.  One has length 38, the other has length 51.  Interestingly, 37 clauses (counting both input and derived clauses) are shared between the two proofs.  Since the first has 38 total clauses---including the empty clause---the second proof can be said to be, in effect, an extension of the first but following a different line of reasoning.


\section{Discussion}\label{sec:discussion}
Limitations of our proof analysis methods should be kept in mind.  The experiments were more or less arbitrarily chosen.  We chose problems that were, first of all, solvable---fairly quickly (less than a minute)---by {\provernine}.  Obviously, if {\provernine} cannot solve a problem, then the kind of proof analysis that is in focus here is, of course, not applicable.

As we saw in Section~\ref{sec:tk} and~\ref{sec:chinese}, the presence of redundant axioms in a problem influences our analysis.  When redundant premises are present, shorter proofs can be found, but the number of proofs can increase, perhaps even tipping the scales so much that infinitely many proofs become available.  In such a situation, we have to impose limits on the search (such as on term complexity, the number of clauses to be kept in the set of support, etc.) and cope with the attendant incompleteness of the analysis (that is, accepting the possibility that there may be proofs ``beyond the horizon'').   If one preprocesses a proof-finding problem by choosing a minimal set of premises, the result is that the length of all proofs might increase (as we saw in Section~\ref{sec:chinese}), but we might be able to achieve greater satisfaction because the overall search might genuinely ``bottom out'' (as we saw in Section~\ref{sec:tk}) by removing various ``rabbit holes'' that {\provernine} might go down.  (Not all rabbit holes can always be eliminated.)
\section{Conclusion}\label{sec:conclusion}
Towards solving the proof identity problem, we have pointed out that a practical (albeit partial) solution to the problem may be staring us in the face, begging for recognition.  If we are dealing with proofs that can be faithfully represented as a finite set of axioms in clause logic, then the door is open (or, to be more modest, may not be closed) to an investigation of the ``space'' of possible proofs (solutions to the proof-finding problem) that Hilbert suggests.

As often happens in theorem proving, totally automated, universally-applicable methods seem to be elusive, but that doesn't mean we are justified in shying away from them.  More specifically, our approach here relies on the ability of a theorem prover to saturate a set of clauses (in effect, to draw all possible inferences from the clauses); but doing so is, in general, theoretically impossible.  We carried out experiments in the TPTP library showing that, in at least some cases, we can practically enumerate all proofs, a result having theoretical interest and which permits concrete applications of ``proof analysis''.



Do\v{z}en concludes his discussion of two natural deduction-based approaches to the proof identity problem by emphasizing that a serious investigation of the issue requires us to shift perspective:

\begin{quotation}
The question we have discussed here [the proof identity problem] suggests a perspective in logic---or perhaps we may say a dimension---that has not been explored enough. Logicians were, and still are, interested mostly in provability, and not in proofs. This is so even in proof theory. When we address the question of identity of proofs we have certainly left the realm of provability, and entered into the realm of proofs.
\end{quotation}

Moving into the realm of proofs (as opposed to staying in the realm of provability) requires, it would seem, a new set of skills.  The question to which we kept returning---are these two proofs materially different?---needs to be tackled afresh every time it is asked.  Even with the heavy lifting being done by an automated theorem prover, we still lack good criteria for going over the finish line and giving a robust answer to the proof identity problem.  Perhaps there are no universally applicable methods.  In Section~\ref{sec:partial}, for instance, we found two distinct minimal-length proofs but argued in the end that they seem to be the same proof.  The argument rested on a symmetry that is built-in to the problem.  The argument is admittedly not conclusive.  Even if one were to be persuaded by such reasoning, one naturally asks: could this symmetry be detected mechanically?

Echoing Do\v{z}en, one might make a parallel remark that in automated theorem proving, at least among those systems that aim to find proofs, the principle focus is derivability, and not the exploration of the space of proofs.  This is an understandable focus, given the sheer difficulty of the proof-finding problem to begin with.  But one can hope that the explorative tradition that one sees in, e.g., {\provernine}, can be extended to currently world-class systems such as {\vampire} or {\eprover}.

For now, we shall merely pose the (rather philosophical) question of whether the Canonicity Principle in Section~\ref{sec:introduction}---``every theorem has a canonical proof''---is compatible with the everyday mathematical truism (and apparently confirmed, in part, by everyday experience with automated theorem proving systems, as well as the results discussed in Section~\ref{sec:experiments}) that some theorems can be proved in at least two ways, even from the (exact) same premises.
\bibliographystyle{splncs03}
\bibliography{proof-analysis}
\end{document}